\documentclass{PoS}

\title{Correlated D-meson decays competing against thermal QGP dilepton radiation}

\ShortTitle{Correlated D-meson decays competing against thermal QGP dilepton radiation }

\author{\speaker{Thomas Lang}\\
        Frankfurt Institute for Advanced Studies (FIAS)\\
        E-mail: \email{lang@th.physik.uni-frankfurt.de}}

\author{Hendrik van Hees\\
        Frankfurt Institute for Advanced Studies (FIAS)\\
         E-mail: \email{hees@th.physik.uni-frankfurt.de}}

\author{Jan Steinheimer\\
        Lawrence Berkeley National Laboratory\\
         E-mail: \email{jsfroschauer@lbl.gov}}

\author{Marcus Bleicher\\
        Frankfurt Institute for Advanced Studies (FIAS)\\
         E-mail: \email{bleicher@th.physik.uni-frankfurt.de}}

\abstract{The QGP that might be created 
in ultrarelativistic heavy-ion collisions is expected to radiate thermal dilepton 
radiation. However, this thermal dilepton radiation interferes with 
dileptons originating from hadron decays. In the invariant mass region 
between the $\phi$ and $J/\psi$ peak ($1\,$GeV$\lesssim M_{\ell^+ \ell^-} \lesssim 3 \,$GeV) 
the most substantial background of 
hadron decays originates from correlated D$\bar{\mathrm{D}}$-meson decays. 
We evaluate this background using a Langevin simulation for charm quarks. 
As background medium we utilize the well-tested UrQMD-hybrid model. 
The required drag and diffusion coefficients are taken from a 
resonance approach. The decoupling of the charm quarks 
from the hot medium is performed 
at a temperature of $130\,$MeV and as hadronization mechanism 
a coalescence approach is chosen. This model 
for charm quark interactions with the medium has already been 
successfully applied to the study of the medium modification and 
the elliptic flow at 
FAIR, RHIC and LHC energies. In this proceeding we present 
our results for the dilepton radiation from correlated D$\bar{\mathrm{D}}$ 
decays at RHIC energy in comparison to PHENIX measurements in the invariant 
mass range between 1 and 3 GeV using different interaction scenarios. 
These results can be utilized to estimate the thermal QGP radiation. }

\FullConference{8th International Workshop on Critical Point and Onset of
Deconfinement\\
                 March 11-15\\
                 Napa, California, USA}

\begin{document}
{
\section{Introduction and model description}

Charm quarks are an excellent probe for the exploration of 
the medium created in ultrarelativistic heavy-ion collisions. 
Due to their high mass well above the medium temperature 
they are only produced in hard parton-parton collisions in the 
pre-equilibrium phase of heavy-ion collisions. 
In the following the charm quarks interact with the evolving medium and 
hadronize to D- and $\bar{\mathrm{D}}$-mesons. The decay products of these D$\bar{\mathrm{D}}$-mesons 
can finally be measured and help to shed light on the medium properties and the interaction 
of the D-mesons with the medium. 

Moreover, D-mesons can be utilized to draw conclusions about the thermal QGP radiation. 
The correlated dileptons from D$\bar{\mathrm{D}}$-meson decays account for the overwhelming  
contribution of the hadron decay background in the 
invariant-mass range between the $\phi$ and $J/\psi$ peak of
approximately $1\,\mathrm{GeV}$ to $3\,\mathrm{GeV}$. 
The dilepton radiation not originating from hadron decays can be accounted to 
thermal QGP radiation. 
Since the invariant mass spectrum of dilepton radiation at RHIC energy 
has been measured by PHENIX, the estimation of the hadron-decay background leads 
in turn to an estimation of the thermal QGP radiation. \\

In this letter we explore the invariant mass spectrum of dileptons 
from correlated D$\bar{\mathrm{D}}$-meson decays in Au+Au collisions at 
$\sqrt{s_{NN}}=200\,\mathrm{GeV}$. 
Since charm quarks are heavy particles, 
we can use a relativistic Langevin approach \cite{vanHees:2007me,Rapp:2009my,He:2011yi} for the propagation 
of charm quarks in the medium consisting of light particles. 
The required drag and diffusion coefficients are taken from a resonance approach \cite{vanHees:2005wb}, 
where the existence of D-meson resonances in the QGP is assumed. 
As background medium for this approach 
we utilize the UrQMD-hybrid model \cite{Petersen:2008dd}. This model provides us 
with a realistic, well-tested medium evolution including event-by-event fluctuations. 

The space-time production coordinates of the charm quarks are assigned 
based on a time-resolved ``Glauber'' approach utilizing the UrQMD model \cite{Bass:1999tu,Bleicher:1999xi}. 
We perform a first UrQMD run excluding interactions between the colliding nuclei and save the nucleon-nucleon space-time coordinates. 
These coordinates are used in a second, full UrQMD run as (possible) production space-time coordinates for the charm quarks. 

For the initial charm-quark momenta at RHIC energy we use \cite{vanHees:2007me,Lang:2012cx} 
\begin{equation}
\frac{\mathrm{d}N}{\mathrm{d}p_T}=\frac{C\cdot\left(1 + A_1\cdot p_T^2\right)^2\cdot p_T}{\left(1+A_2\cdot p_T^2\right)^{A_3}},
\end{equation}
with $A_1=2.0/\mathrm{GeV}^2$, $A_2=0.1471/\mathrm{GeV}^2$, $A_3=21.0$. 
$C$ is an arbitrary normalization constant with the unit $1/\mathrm{GeV}^2$. 
Due to conservation laws the charm and anticharm quarks are produced in pairs and are 
emitted back-to-back. 

Starting with these initial charm-quark distributions we perform the Langevin 
calculation in the hydrodynamic state of the medium evolution (cf.\ \cite{Lang:2012cx} for details). 

The charm quarks decouple from the hot medium at a temperature of $130\,$MeV and their 
hadronization is performed employing a coalescence mechanism \cite{Lang:2012cx}. 
The introduced Langevin model 
for charm-quark propagation has been successfully applied to the calculation  
of the medium modification factor $R_{AA}$ and the elliptic flow $v_2$ 
at FAIR, RHIC and LHC energies \cite{Lang:2012cx,Lang:2012yf,Lang:2012vv,Lang:2013cca}.

\section{The invariant mass spectra of D-meson decays}

For the comparison of our approach to the invariant mass spectrum of dielectron pairs 
measured by PHENIX we assume three different thermalization scenarios: \\
No interaction: The charm quarks do not interact with the medium. 
However, their initial back-to-back correlation is modified due to 
their hadronization via the coalescence mechanism. \\
Physical thermalization: We use the physical drag and diffusion coefficients 
from the resonance model. Utilizing these coefficients a good agreement to data for the medium 
modification and elliptic flow of heavy quarks at RHIC and LHC has been reached \cite{Lang:2012cx,Lang:2012yf,Lang:2012vv}. \\ 
Extreme thermalization: The drag and diffusion coefficients are multiplied by a factor of 20 and a 
nearly full thermalization of the charm quarks to the background medium is achieved. \\

Our results for a minimum bias calculation ($\sigma/\sigma_{\mathrm{tot}}=$0\%-92\%) 
with the appropriate experimental cuts are shown in Figure \ref{RHICdata} in 
comparison to the PHENIX measurements. 
\begin{figure}[h!]
\center
\includegraphics[width=0.75\textwidth]{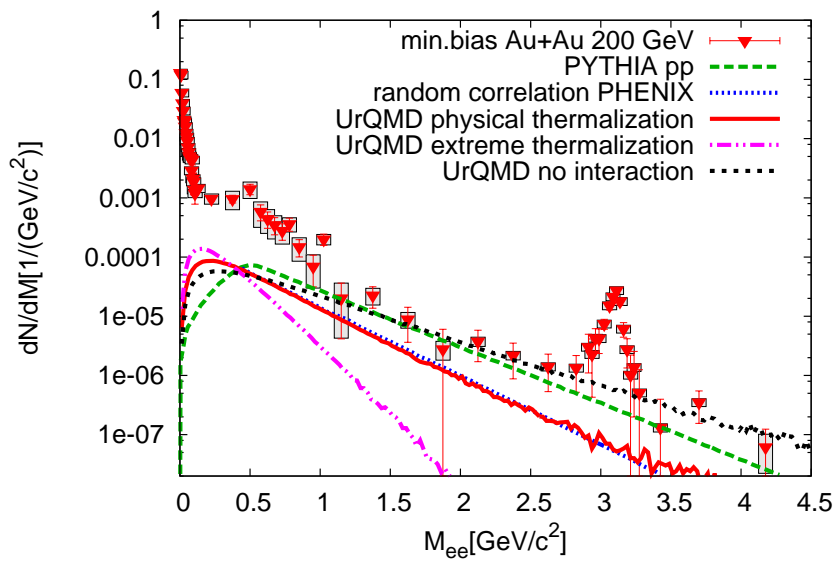}
\caption{(Color online) Invariant mass spectra of electrons in Au+Au
  collisions at $\sqrt{s}_{NN}=200\,\mathrm{GeV}$ in the centrality range
  of 0-92\% (min. bias). A rapidity cut of $|y|<0.35$ and a momentum cut
  of $p_T^e=0.2\mathrm{GeV/c}$ are applied. The dilepton data points are
  taken from a PHENIX measurement \cite{Adare:2009qk}. They are compared
  to calculations of electrons from D-meson decays. Also the pp
  calculation in PYTHIA and the random correlation calculation are taken
  from PHENIX \cite{Adare:2009qk}. The UrQMD calculations show three
  different scenarios: no interaction with the medium, physical drag and
  diffusion coefficients and extreme thermalization with 20 times higher
  coefficients. The difference between our physical scenario and the
  measured dilepton decays might be due to thermal radiation from the
  medium. The yields are normalized to the PYTHIA pp yield taken from \cite{Adare:2009qk}.}
\label{RHICdata}
\end{figure}

The scenario without any medium interactions matches the data in the invariant mass range 
between $1$ and $3\,\mathrm{GeV}$ quite well. The calculation utilizing physical drag and 
diffusion coefficients results in a softer invariant mass spectrum. This leads to 
an underestimation of the data. This underestimation grows significantly with the 
invariant mass. The difference between this calculation and the 
data might be caused by thermal QGP radiation, which is expected to dominate among all 
other thermal sources in this invariant mass range. For the scenario with 20-fold enhanced 
drag and diffusion coefficients a by far softer spectrum is produced and therefore the gap 
between our calculation and the measurements further increases. 

The difference between our calculation neglecting interactions and the PYTHIA pp calculations 
stems from the differing initial charm-quark distribution and the inclusion of coalescence 
in our calculation. The discrepancy between our calculation assuming extreme thermalization 
and the PHENIX random correlation is caused by different assumptions for the random correlation. 
While PHENIX assumes a flat angular correlation, we assume a 3-dimensional random 
correlation function which corresponds to the geometrical $\Delta\Phi$ dependence (cf.\ \cite{Lang:2013wya} for details). \\

In summary, we have presented an estimation of the hadronic dielectron background 
originating in correlated D$\bar{\mathrm{D}}$-meson decay in the invariant mass region between 
$1$ and $3\,\mathrm{GeV}$. Our physical calculation suggests a significant contribution 
of thermal QGP radiation in this invariant mass range which might serve  
as a signal for the creation of a QGP.

}

\end{document}